\documentclass[graybox]{svmult}

\usepackage{mathptmx}       
\usepackage{helvet}         
\usepackage{courier}        
\usepackage{type1cm}        
\usepackage{makeidx}         
\usepackage{graphicx}        
\usepackage{sidecap}
\usepackage{subfigure}
\usepackage{multicol}        
\usepackage[bottom]{footmisc}

\usepackage{natbib}         

\newcommand{\aap}{A\&A}
\newcommand{\apj}{ApJ}
\newcommand{\aj}{AJ}
\newcommand{\apjl}{ApJL}
\newcommand{\apjs}{ApJS}
\newcommand{\mnras}{MNRAS}
\newcommand{\nat}{Nature}

\makeindex             

\begin{document}

\title*{Boxy/peanut/X bulges, barlenses and the thick part of galactic
  bars: What are they and how did they form?} 
 \titlerunning{The thick part of galactic bars} 
 \author{E. Athanassoula}
\institute{E. Athanassoula \at Aix Marseille Universit\'e, CNRS, LAM (Laboratoire
d'Astrophysique de Marseille),       
UMR 7326, 13388 Marseille 13, France, \email{lia@lam.fr}}
%
%
\maketitle
\abstract{Bars have a complex three-dimensional shape. In particular
  their inner part is vertically much thicker than the parts further out. Viewed
  edge-on, the thick part of the bar is what is commonly known as
  a boxy-, peanut- or 
  X- bulge and viewed face-on it is referred to as a barlens. These 
  components are due to disc and bar instabilities and are
  composed of disc material.
  I review here their formation,
  evolution and dynamics, using simulations, orbital
  structure theory and comparisons to observations.  
}  

\section{Introduction}
\label{sec:intro}

Boxy/peanut/X (for short B/P/X, or B/P) bulges protrude out of the
central region of galactic 
discs viewed edge-on. Their name comes from their shape, which is
reminiscent of a box, a peanut or an `X' structure. Good examples are
NGC 1381 and ESO 151-G004. There have been many observational studies
of such objects over the years, while their formation and evolution have
also been extensively studied with the help of simulations, both to understand
their origin and as a link to secular evolution. Orbital studies
have provided candidate families for the backbone of this structure. 

Barlens components (bl for short) were introduced into the picture only quite
recently 
\citep{Laurikainen.SBK.11}. They are defined as
``lens-like structures embedded in the bars'' \citep{Laurikainen.SABBJ.13}.
They are thus found in the central part of barred galaxies
``but are generally distinct from nuclear lenses by their much larger
sizes'' \citep{Laurikainen.SBK.11}. They are also distinct from standard lenses
\citep{Kormendy.79} because they are shorter than bars 
\citep{Laurikainen.SABBJ.13} and because on the bar major axis
they blend smoothly in the bar radial density profile, without
having any steep drop \citep{Laurikainen.SABHE.14, Athanassoula.LSB.14}. 
NGC 4314 and NGC 4608 are good examples of galaxies with a barlens
component. Images of further example galaxies can be found in 
the NIRS0S (Near Infrared S0 survey) atlas
\citep{Laurikainen.SBK.11}, the Hubble atlas \citep{Sandage.61}, the
S$^4$G (Spitzer Stellar Structure Survey of Galaxies) sample
\citep{Sheth.P.10}, 
as well as in Fig. 2 of \cite{Buta.LSBK.06} and Fig. 8 
and 12 of \cite{Gadotti.08}. 

Here I will discuss how these components form and evolve and what their
properties and dynamics are, basing this discussion on
simulations, orbital structure results and on comparison with
observations. I will discuss neither the Milky Way bulge, nor bulges
in a cosmological setting, and will not give a full account of
observations, since all three subjects will be covered elsewhere in
this book. I first review  
orbital structure results (Sect.~\ref{sec:orbits}), focusing on the
families that can be building blocks  of B/P/X/bl structures. I then
turn to simulation results (Sect.~\ref{sec:sims}). In particular, in 
Sect.~\ref{subsec:3Dbars} I discuss
the ensuing shape of bars and the B/P extent. Comparison with
observations is the subject of Sect.~\ref{sec:observations}:
morphology and photometry in Sect.~\ref{subsec:MP}; kinematics in
Sect.~\ref{subsec:kinematics}. I discuss theoretical aspects of the
barlens component in Sect.~\ref{sec:barlens}.
Recent reviews on this or related subjects have been given by
\cite{Kormendy.Kennicutt.04}, \cite{Athanassoula.08, Athanassoula.13a} and
\cite{Kormendy.08, Kormendy.13}. 
 
\section{Orbital structure}
\label{sec:orbits}

In order to understand the structure, kinematics or dynamics of a given
galaxy, or of any of its substructures, it is necessary first to
understand the orbits that constitute it. Particularly important for
this are the periodic orbits -- i.e. orbits that close in a given
reference 
frame after a number of rotations -- which constitute the 
backbone of the structure. These
come in two types. {\it Stable periodic orbits} trap around them
regular orbits, while {\it unstable periodic orbits} are linked to chaos. 
The latter, however, can also, in certain cases, contribute to the outline of 
structures.

\subsection{Periodic orbits in two dimensions}
\label{subsec:2Dpo}

The orbital structure of bars in two dimensions (2D) is relatively simple.
The main backbone here is the $x_1$ family, constituted of orbits which, in
a frame of reference co-rotating with the bar, 
close after two radial oscillations and one revolution around the 
centre, i.e. are in 2:1 resonance 
\citep{Contopoulos.Papayannopoulos.80, Athanassoula.BMP.83}. 
They are elongated along the bar and their axial ratio varies
with distance from the centre, but also from one model to another. At
their apocentres they often have cusps or loops 
(see \citealt{Athanassoula.92a}
for a study of their morphology). There are other families of
orbits, such as the $x_2$ -- which is also 2:1 but is elongated
perpendicular 
to the bar -- the 3:1, or the 4:1, but they are less important for the
global bar structure, although they are related to several specific
aspects such as the shape of the bar, or the structure of the inner kpc. 

In 2D studies, by construction, we can study orbital stability only in
the plane and the trapped
orbits are also planar. A fair fraction of the $x_1$ orbits are stable,
but the amount of chaos depends strongly on the properties of the bar,
such as its mass, axial ratio etc. \citep[e.g.][]{Athanassoula.BMP.83,
  Manos.Athanassoula.11}. 

\subsection{Periodic orbits in three dimensions}
\label{subsec:3Dpo}

\begin{figure*}
\centering
\includegraphics[scale=0.30, angle=0.0]{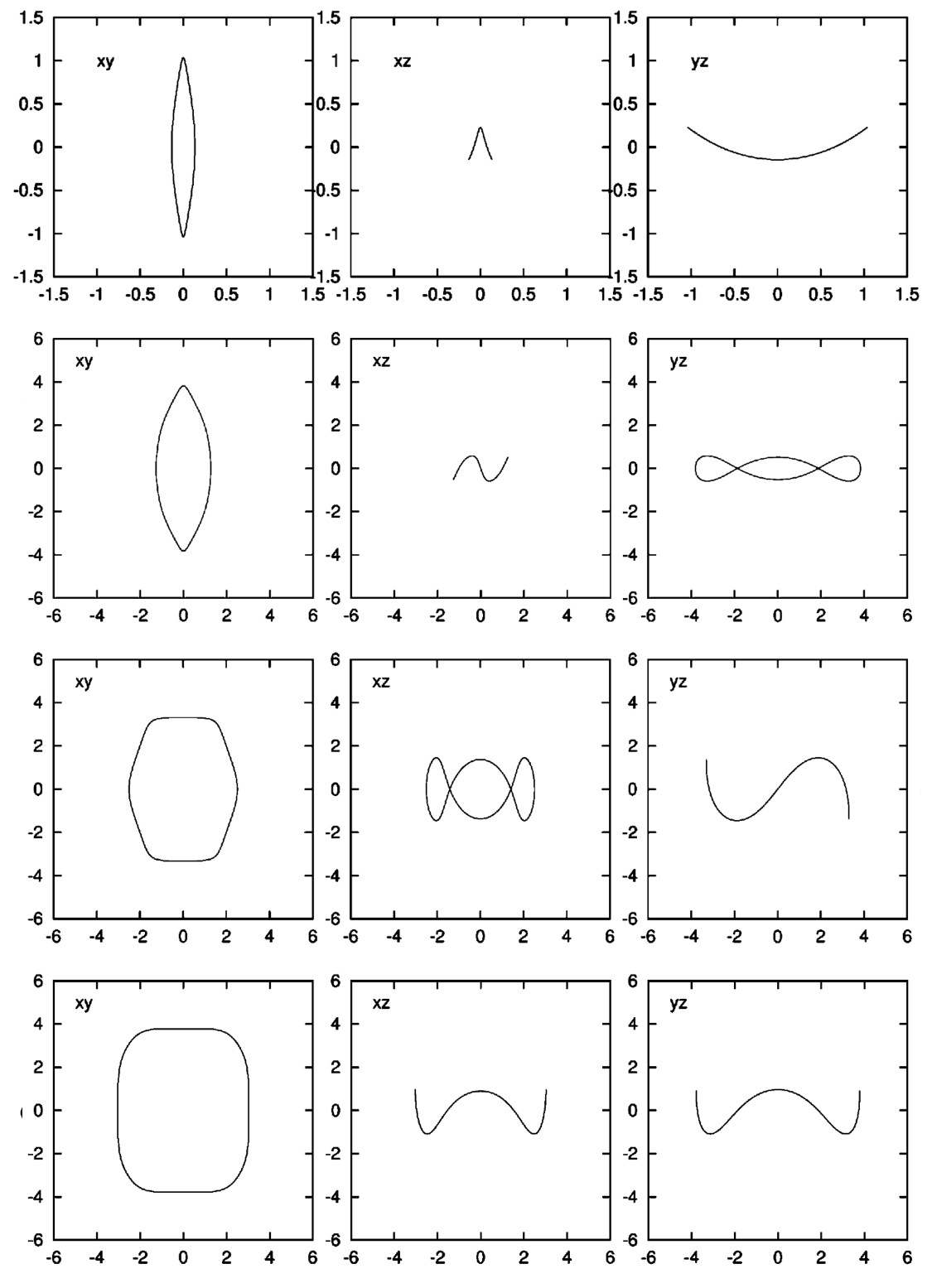}
\caption{ 
Four examples of periodic orbits from vertical families. From top to bottom
we have members of the $x_1v_1$, $x_1v_3$, $x_1v_4$ and $x_1v_5$ families.
There are three views for each orbit: face-on (left), end-on (middle)
and side-on (right). Note that the linear
scales used in the plots of the various orbits are not the same.
(This plot is a composite made of parts of figures 8, 9, 10 and 14 of
Skokos, Patsis \& 
Athanassoula, {\it Orbital dynamics of three-dimensional bars - I. The
backbone of three-dimensional bars. A fiducial case}, 2002, MNRAS, 333,
847.)
}
\label{fig:4orb3d}
\end{figure*}

In three dimensions (3D) the orbital structure becomes much more
complex, even for planar orbits \citep[e.g.][]{Pfenniger.84,
Skokos.PA.02a, Skokos.PA.02b, Harsoula.Kalapotharakos.09, 
Patsis.Katsanikas.14a, Patsis.Katsanikas.14b}. 
Indeed -- while in two dimensions periodic orbits can be  
stable or unstable depending on their response to in-plane
perturbations  -- in 3D all orbits, including the planar periodic ones, can be subject
to vertical perturbations, which in turn can introduce instability in
the system. The latter is particularly important
for our subject matter. At the energy value where a family turns
from stable to unstable a new stable family is generated by
bifurcation, and this may play an important role in the dynamics of the
system. Thus, in barred galaxies there are a number of vertical families
named by Skokos et al. as as $x_1v_1$, $x_1v_2$, $x_1v_3$
etc.\footnote{For certain 
  potentials there is also the z3.1s family whose 
  morphology resembles that of the $x_1v_4$ family, but it is 
  not related to the $x_1$ tree. Another potentially useful family is
  the x1mul2 \citep{Patsis.Katsanikas.14a}.}. These bifurcate
from the $x_1$ family at the main vertical resonances, such as
the 2:1, 3:1, 4:1 
etc. There are generally two per resonance, one which crosses the
symmetry plane perpendicular to the bar major axis at $z$ = 0 and the
other with $\dot{z}$ = 0. Hence, $x_1v_1$ and $x_1v_2$ correspond to
the 2:1 vertical resonance,
$x_1v_3$ and $x_1v_4$ to the 3:1 etc. Trapping around these families determines the
vertical thickness and structure of the bar. These vertical
families, together with the $x_1$ family  
from which they bifurcate, form what is often referred to as the $x_1$
tree \citep*{Skokos.PA.02a}. Thus the backbone of a 3D bar is not the $x_1$
family but the $x_1$ tree. Examples of members of the main four vertical
families are given in Fig.~\ref{fig:4orb3d}. These plots
are taken from the work of \cite*{Skokos.PA.02a}
where the bar is along the $y$ axis, so that the end-on view\footnote{In 
the end-on view the galaxy is observed edge-on with the line of sight
along the bar major axis.} is the projection on the ($x$, $z$) plane
and the side-on view\footnote{In 
the side-on view the galaxy is observed edge-on with the line of sight
perpendicular to the bar major axis.} is the projection on the ($y$,
$z$) plane. Since the bar potential is symmetric with respect
to the equatorial plane, for each   
periodic orbit there is also its corresponding symmetric one (not
shown here). Thus the side-on view of an $x_1v_1$ orbit has the shape of either a
smile ($\smile$) or a frown ($\frown$). 

The orbits of each family can be useful for building a vertically thick
structure only within certain energy ranges and fill only specific regions of the 3D
space. As a result, these orbits
do not thicken vertically the whole bar; instead they only thicken
its inner parts. Thus, if the vertical families are
populated, the bar should have two
parts: an inner one that is vertically thick and an outer one
which is vertically thin. 

A further point to note is that the members of the various
families have different extents along the bar 
major axis relative to the bar length, and their vertical extent
and face-on shapes differ considerably from one family to
another. The $x_1v_1$ family, 
which bifurcates at the lowest energy, has the smallest extent along
the bar major axis and the largest extent perpendicular to the equatorial
plane. Its face-on shape is rather elongated. Higher order families,
compared to lower order ones,
bifurcate at higher energies, have a larger extent along the bar major
axis close to the equatorial plane and a smaller one perpendicular
to this plane. Their face-on outline is much less
elongated. Measuring the ratio of the bar length (as 
determined from the orbits that constitute it) to the length of the
thick part (also from its orbits) (\cite{Patsis.SA.02}, hereafter
PSA02) find that this  
number for the $x_1v_1$ is roughly in the range [2., 4.], while for the 
$x_1v_4$ family it is around 1.1 to 1.3. A note of caution is necessary
though: the only realistic bar  potential used so far in 3D orbital
calculations is the Ferrers' bar potential \citep{Ferrers.77}. It is
thus not possible to  
check to what extent these numbers are model dependent. Moreover, the
density corresponding to the Ferrers potential does not have an
appropriate side-on shape, i.e. it is neither boxy- nor peanut- shaped. 
Performing more orbital structure calculations using a yet more
realistic potential would be highly desirable at this stage.
  
\subsection{The role of chaotic orbits}
\label{subsec:chaos}

Stable periodic orbits and regular orbits trapped around them are not
the only way of building the galactic structures we are discussing
here. Unstable periodic orbits are linked to chaos and could, in some
cases, provide an alternative. Indeed sticky chaotic orbits may also
contribute to such structures either if they stick to regular tori
around the stable families or to unstable asymptotic
curves of the unstable periodic orbits \citep{Contopoulos.Harsoula.08}.

\cite{Patsis.Katsanikas.14a, Patsis.Katsanikas.14b} examined the
evolution of the phase space in a 3D bar and underlined the role that
chaotic phenomena may play in building the B/P/X structures. This is a
promising alternative and merits further work to establish its role
in galaxies. High quality N-body simulations, provided they are
realistic, are a perfect test bed for such types of studies, because
they offer not only the possibility of viewing the structures from any
desired viewing angles, but also allow studies at the level of
individual orbits. They can thus give information on the amount of
chaotic orbits and also on their specific contributions to the B/P/X
structures. First steps in this direction have already been made
\citep{Athanassoula.05c, Harsoula.Kalapotharakos.09, Manos.Machado.14}
and more specific applications are underway. 
 
\section{Simulations}
\label{sec:sims}

\subsection{General description}
\label{subsec:general}

Although inklings of a boxy/peanut structure can be already seen in the
edge-on views of the simulations of \cite{Hohl.Zang.79} and
\cite{Miller.Smith.79}, the first to show it convincingly
were those of \cite{Combes.Sanders.81}. They were followed by 
\cite{Combes.DFP.90}, who found
a B/P morphology in all their bar-forming simulations 
viewed side-on. 
This forms somewhat after the bar, with a delay of the order of a Gyr
(see Sect.~\ref{subsec:tevol}). Using axisymmetric definitions for the
resonances, the authors {found that the horizontal and vertical inner
  Lindblad resonances (ILRs)
  coincide by the end of the simulations. This, however, is presumably
  model dependent \citep[see e.g.][for a different
    behaviour]{Quillen.MSQDM.14}}.
Both \cite{Combes.DFP.90} and \cite{Pfenniger.Friedli.91} found
that the backbone of the peanut should be a vertically 2:1 family. 

\cite{Pfenniger.Friedli.91} and \cite{Raha.SJK.91}, running similar
simulations, found that the 
formation of the B/P structure is preceded by an asymmetric
phase during which the equatorial plane is not a symmetry plane anymore.
In the latter of these two papers this was ascribed to the
fire-hose instability \citep{Toomre.66}, while in the former to orbital
instabilities and the ensuing families (Sect.~\ref{sec:orbits}). 
Both \cite{Combes.DFP.90} and \cite{Raha.SJK.91} note that the B/P
formation is associated with a drop of the bar
strength, which in many cases can be strong and sharp. The latter work
conjectured that ``bars may be [...] even destroyed by this
instability''.   
However, \cite{Debattista.CMM.04, Debattista.MCMWQ.06} ran a larger
set of simulations and found no clear case of bar destruction. From my
own, yet larger set of simulations, I also found the same result
(unpublished). Strictly speaking, this does not prove that a bar
destruction can not occur, it just shows that it is rather unlikely,
unless this occurs in a part of the parameter space which
has not yet been explored. Note also that after its
sharp decrease, the bar strength starts increasing again, or at least
stays relatively constant (see Sect.~\ref{subsec:tevol}).

\citet[][hereafter AM02]{Athanassoula.Misiriotis.02} and 
\cite{Athanassoula.05a} 
`observed' the bars and B/P bulges in their simulations and obtained
specific results on their shape, extent and kinematics.
The quantitative estimates they obtained showed
clearly that the B/P bulges are shorter than bars. 
AM02 and \cite{Athanassoula.03, Athanassoula.05a} 
showed that stronger bars produced on average stronger B/P 
bulges, i.e. bulges which extended further out from the galactic
plane, thus confirming the observational result of
\cite{Lutticke.DP.00b}.  
They also found 
that the thick part of relatively weak bars generally
has a boxy shape, that of 
stronger ones a peanut shape, and the very strong ones an X shape. 

\cite{Mihos.WHMB.95} simulated a minor merger of a disc galaxy with
its satellite. This induces a strong bar which
forms a clear X shape. Strictly speaking, this is not really an example
of a B/P formation from a merger, since all the companion does is to
drive a bar, which, once formed, buckles and thickens
vertically. This driven bar is very strong and according to the
results discussed above, one would expect an `X' shaped bulge to form,
as indeed occurred in the simulation.    

\begin{figure*}
\centering
\includegraphics[scale=0.30, angle=0.0]{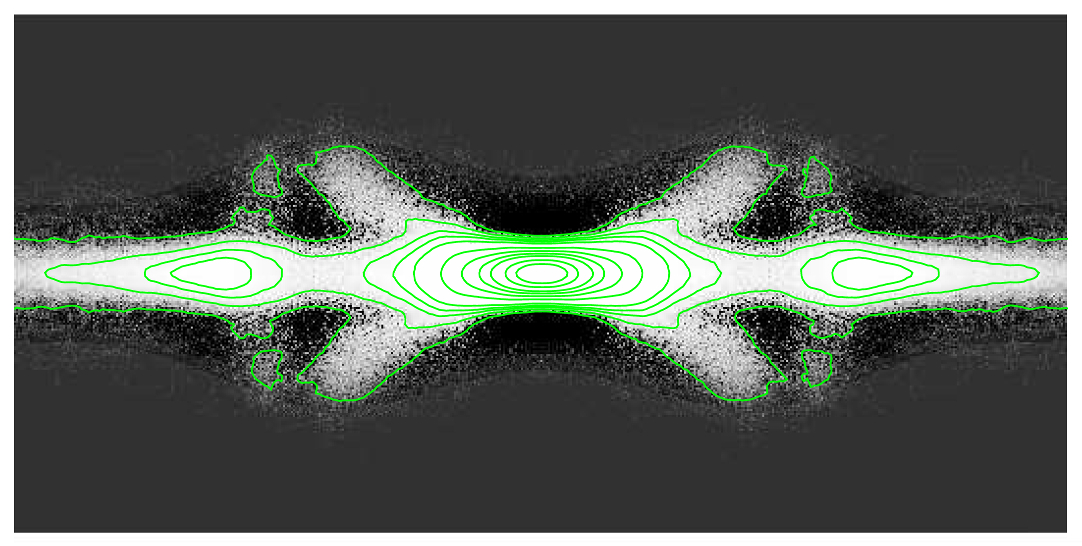}
\caption{Unsharp masked image of a simulation with a strong bar
  viewed side-on. The lighter shades in the grey scale plot
  correspond to higher values and the darkest areas
  correspond to negative values. The isodensities were
  chosen so as to show best the relevant features. 
(From figure 6 of Athanassoula, {\it On the nature of bulges in general and
  of box/peanut bulges in particular: input from N-body simulations},
  2005, MNRAS, 358, 1477.) 
}
\label{fig:unsharp-mask}
\end{figure*}

\cite{Athanassoula.05a} unsharp masked\footnote{Unsharp masking, also
  called median filtering, consists in replacing the value of each
  pixel by the difference between it and the median of all pixel
  values within a circular aperture centred on the pixel. This
  highlights sharp features.} a number of images from N-body
simulations with the disc viewed edge-on and found a number of interesting
morphological features. An example is given in
Fig.~\ref{fig:unsharp-mask}, which clearly shows an X-shape, whose four
arms do not meet at the centre but in pairs at a distance from
it. In other examples though (not shown here) these four arms meet
together in the centre. Schematically, these two types of X shapes can
be shown as $>$--$<$ and $><$, respectively. They were later found
also in observations and were dubbed off-centred and centred Xs,
respectively \citep{Bureau.AADBF.06}. There are also 
two clear maxima, one on either side of the centre of the simulated
galaxy. In the example of Fig.~\ref{fig:unsharp-mask} they are due to
an inner ring, but they could also have been due to a superposition of
appropriate orbit families (PSA02).

\cite{Martinez.VSH.06} witnessed in her simulation a second buckling
event occurring 
between 5 and 8 Gyr, i.e. when the bar is in its secular evolution
phase. At the beginning of this time range, its length is already
of the order of 12 $\pm$ 1 kpc, and continues growing 
after the end of the second buckling, reaching roughly 16
kpc at 12 Gyr. During the first buckling, the asymmetry is strongest in the
region closer to the centre and during the second
one roughly in  the middle of the bar region. Such events can also be
seen, or inferred, in other simulations 
\cite[e.g.][]{ONeil.Dubinski.03, Athanassoula.05b, 
  Athanassoula.MR.13} and 
even a triple buckling has been reported \citep{Debattista.MCMWQ.06}.  

Including gas in simulations may or may not suppress buckling
\footnote {Peanut formation without buckling has also been found in
  simulations with no gas \citep{Quillen.MSQDM.14}.}. 
\cite{Berentzen.HSF.98, Berentzen.SMVH.07} and  
  \cite{Villa.VSH.10} use an isothermal gas and 
note that the vertical buckling is much less pronounced than in a
similar but collisionless simulation, to the point of being difficult
to detect by simple visual inspection. They find that with increasing
gas fraction, both the buckling and the B/P strength decrease. 
When radiative cooling is included, buckling 
is prohibited \citep{Debattista.MCMWQ.06,
  Wozniak.Dansac.09}. This is in agreement with the bar strength 
evolution of the simulations in \cite{Athanassoula.MR.13}, coupled to
their peanut strength shown in (\citealt{Iannuzzi.Athanassoula.15}, 
hereafter IA15).

\subsection{Evolution of bar related quantities}
\label{subsec:tevol}
\begin{figure*}[ht!]
%
        \qquad \qquad \qquad \qquad \qquad 
        \subfigure{%
           \label{fig:second}
           \includegraphics[width=0.45\textwidth]{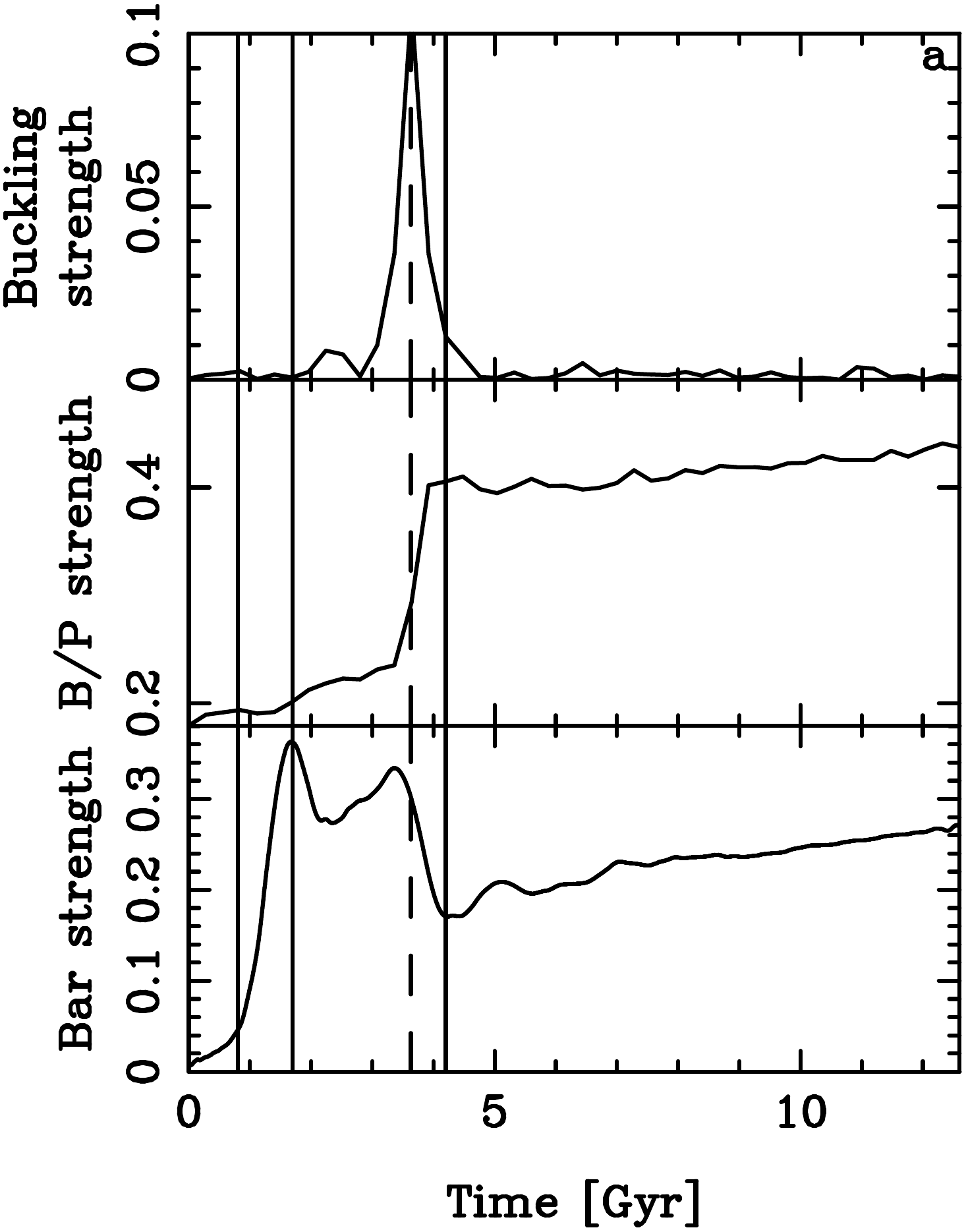}
        }\\ 
        \quad
        \subfigure{%
            \label{fig:third}
            \includegraphics[width=0.45\textwidth]{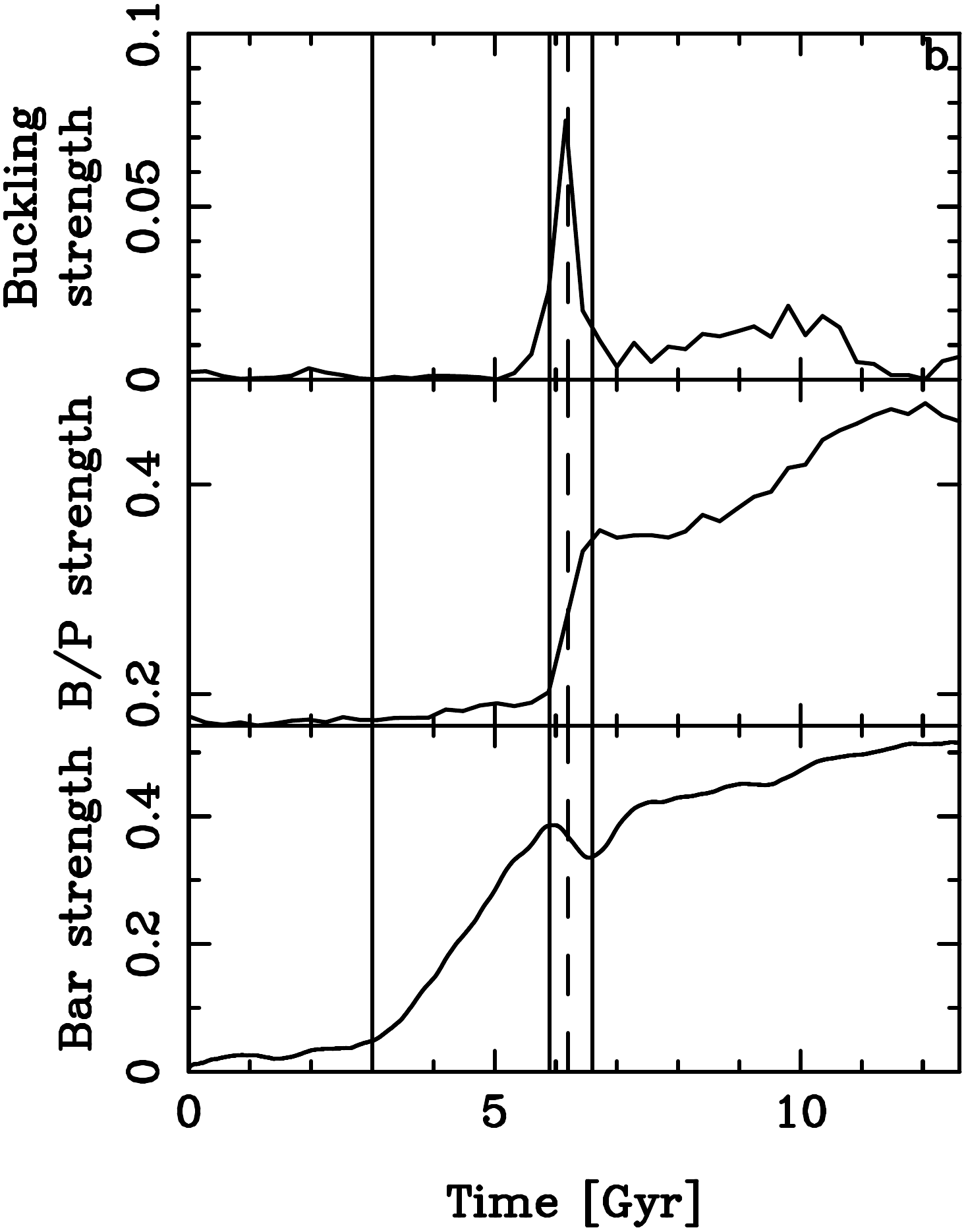}
        }%
        \qquad
        \subfigure{%
           \label{fig:fourth}
           \includegraphics[width=0.45\textwidth]{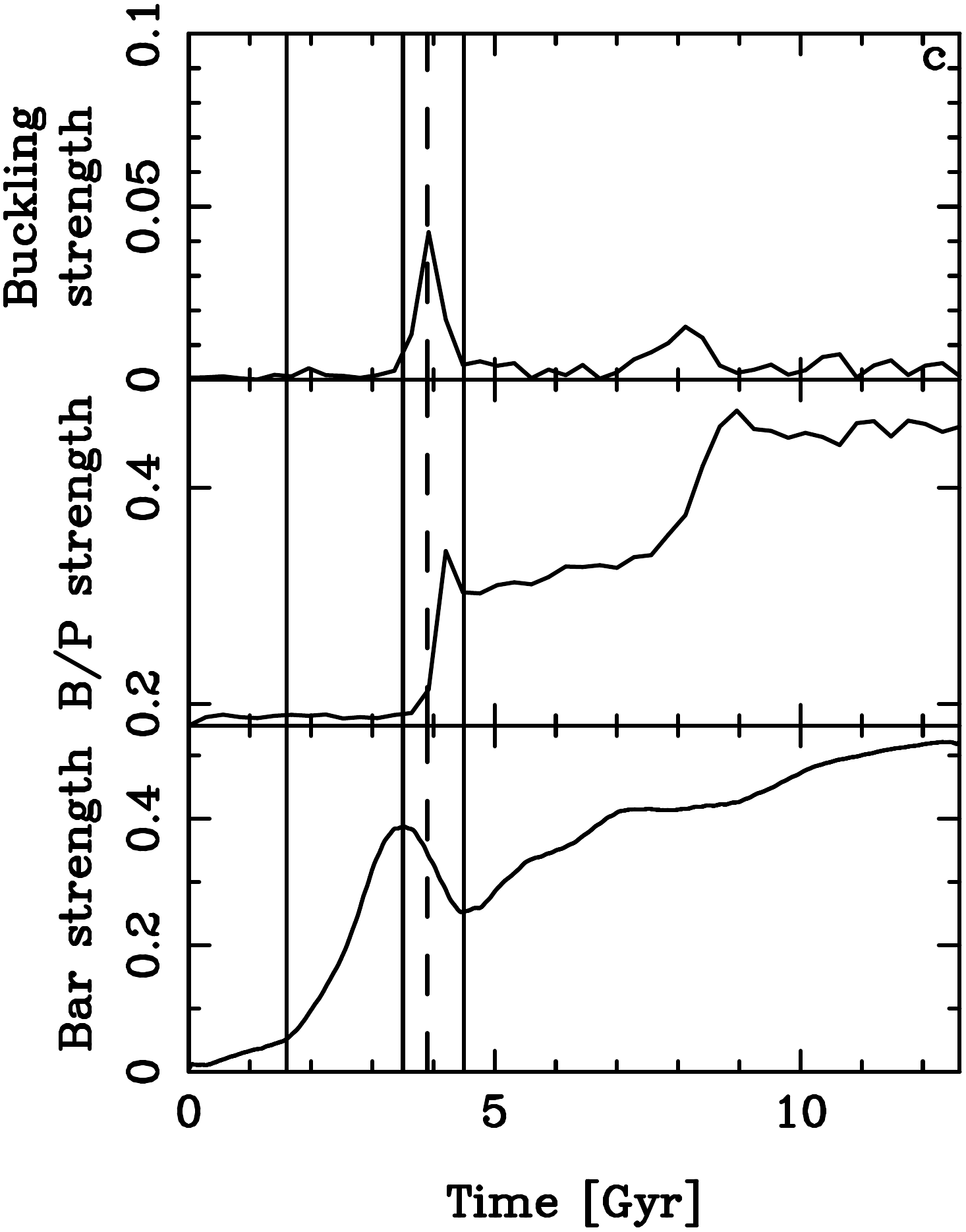}
        }%
%
    \caption{%
        Evolution of bar related quantities for three
        simulations (see text). Each panel has three sub-panels, 
        showing the evolution of 
the asymmetry (top), of the B/P strength (middle) and of the bar
strength as given by the m=2 Fourier component of the density
(bottom). In all panels, the first vertical solid line corresponds to
the time the bar starts growing and the second to the end of the growth
phase (when the amplitude of the bar strength has reached a local
maximum). The third (dashed) vertical line shows the maximum of the
buckling strength (maximum asymmetry) and the last (solid) line the
minimum of the bar strength before the start of the secular evolution phase.
}
\label{fig:tevol}
\end{figure*}

In Fig.~\ref{fig:tevol} I show the time evolution of bar-related
quantities for three different collisionless simulations. The upper panel (a)
corresponds to what is referred to in AM02 
as an MD model, i.e. a model where the disc dominates in the inner region   
(a maximum disc model). The two lower panels correspond to what is
referred to in AM02 as an MH model. Here the halo and the disc
contributions are comparable in the inner parts and the halo plays a
more prominent role in the angular momentum redistribution within the galaxy. 

In the MD model, the bar starts growing very rapidly, roughly 0.8
Gyr from the beginning of the simulation. Its growth phase lasts also 
less than a Gyr, after which the strength of the m=2 component reaches a 
maximum, due to some extent to a strong, but short-lived two-armed spiral 
\citep{Athanassoula.13b}. Between t=3.6 and 4.2 Gyr the bar strength
decreases very strongly and 
rapidly, after which it starts increasing again due to secular
evolution. The buckling strength is measured from the asymmetry with
respect to the equatorial plane and shows a strong and narrow peak at
the buckling time (t=3.65 Gyr). The strength of the B/P increases abruptly
in the time interval during which the bar strength drops. Note
that the time of maximum asymmetry is within this time range. 

The corresponding plots for the first MH run (bottom left panel, b) show the
same qualitative behaviour, but with clear quantitative differences. Namely,
the bar starts growing considerably later (t=3 Gyr), grows more
during the secular evolution phase and reaches a higher strength by
the end of the run. These differences can be easily understood 
because the halo delays bar formation initially, but at later times
helps the bar grow by absorbing angular momentum emitted from the bar
region \citep{Athanassoula.02, Athanassoula.03}.

The third set of plots (bottom right panel, c) is also for an MH run but shows
an interesting difference from the previous simulation, namely there
is a second buckling event, occurring roughly
between 7 and 9 Gyr. The second asymmetry peak is less high and
also broader, i.e. the buckling lasts considerably longer but is less strong. The
time range during which the peanut strength increases is also somewhat
longer, and the increase in B/P strength considerable. 
During that time the bar strength stops increasing and stays
roughly constant.   

The buckling episode can also be accompanied by an abrupt change of
$\sigma_z/\sigma_r$, where $\sigma_z$ and $\sigma_r$ are the $z$ and
radial components of the velocity dispersion, respectively
\citep[e.g.][]{Debattista.MCMWQ.06, Martinez.VSH.06, Athanassoula.08}. 
\cite{Saha.PT.13},
however, present a case where a sharp drop occurs well before the
buckling and propose an alternative indicator, namely the tilt of the
velocity ellipsoid in the meridional plane.               
 
\subsection{The 3D shape of bars}
\label{subsec:3Dbars}

\begin{figure*}
\centering
\includegraphics[scale=0.50, angle=0.0]{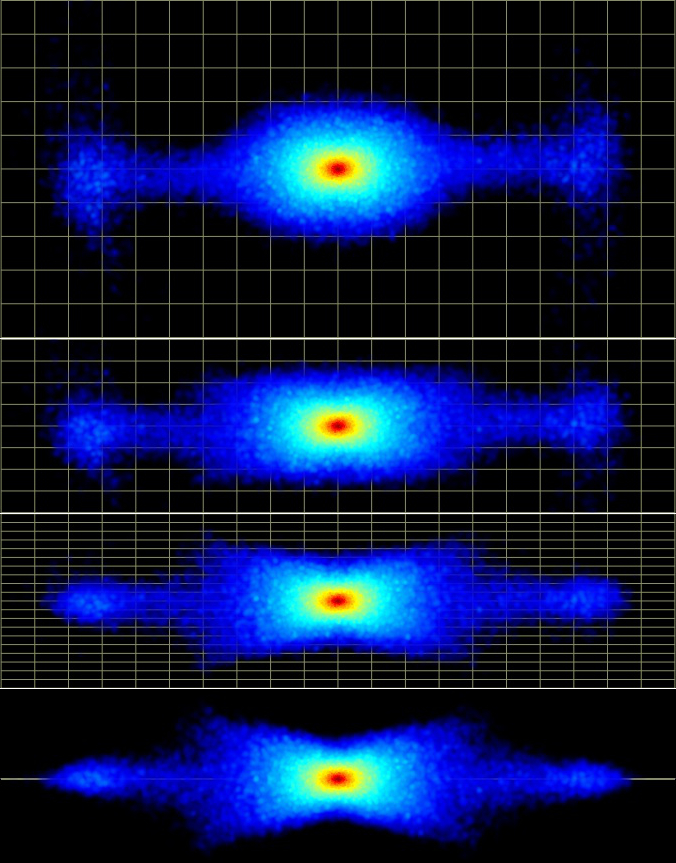}
\caption{Four views of a strong bar 
  (see text). A Cartesian grid with 1 kpc x 1 kpc cell size and
  located on the $z$=0 (equatorial) plane is also shown in all  
  panels, to give a better understanding of perspective and size.   
}
\label{fig:four-views-strong}
\end{figure*}

\cite{Athanassoula.05a} presented evidence on the 3D shape of
bars coming from various sources, including orbital structure
calculations, simulations and 
various observations of real galaxies. All converges to the same
conclusion: Bars have a complex 3D shape with a vertically thick inner
part and a thin outer part. Therefore, a B/P/X shaped component is a
part of a bar, 
and, more specifically, its thick part. To visualise this best, it
is informative to take a snapshot from an N-body simulation of a
bar-forming disc with no classical bulge at a time
when both the bar and the B/P feature have formed. Then select only the
particles which in the ($x,y$) view are located roughly within the outer
isodensities of the bar and visualise them from many
viewing angles. Fig.~\ref{fig:four-views-strong} shows an example for such a
result from a simulation which was chosen so as to have a strong bar
with a somewhat X-like edge-on view. The top 
panel shows the face-on view and the bottom one gives the edge-on one,
with two intermediate viewing angles in between (second and third
panels). For a better
visualisation see the complete animation showing the slow rotation
around the bar major axis in 
http://195.221.212.246:4780/dynam/movie/BPreview/BPreview.avi

In the face-on view the bar is seen to have a length of roughly 8 kpc,
while in the near-side-on views the thick part is seen to have an extent of
roughly 4 kpc, i.e. is clearly less extended than the bar. The ratio
of the two extents argues that the main contributor to the B/P/X
feature could be the $x_1v_1$ family.
The face-on shape of the thick part of the bar can be described as a
squashed oval, because 
the parts of the isodensities near the bar minor axis form nearly 
straight lines parallel to the bar major axis. In the second view
(second panel from the top), the 
outermost parts have a clearly rectangular-like outline, which becomes
X-shaped in the fully side-on view (bottom panel).  
 
Both in the face-on and in the side-on views the outer part of the bar
is quite thin. Near its ends, the bar outline becomes more extended in
the face-on view, with a ansae-like shape. 

\begin{figure*}
\centering
\includegraphics[scale=0.50, angle=0.0]{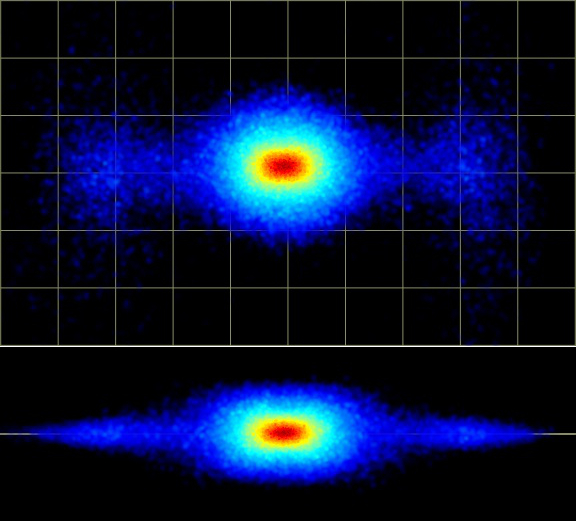}
\caption{Same as in Fig.~\ref{fig:four-views-strong}, but for a weaker
  bar. Only the face-on (upper panel) and side-on (lower
  panel) views are shown. 
}
\label{fig:two-views-weak}
\end{figure*}

Fig.~\ref{fig:two-views-weak} gives similar information, but for a
weaker bar. In the face-on view the bar has a length of roughly
4 kpc, while in the near-side-on views (not shown here) the thick
part is seen to have an extent of roughly 1.5 kpc , i.e. the latter is clearly
less extended than the bar along its major axis. 
The ratio 
of the two extents now argues strongly that the main contributor to the B/P
feature is the $x_1v_1$ family. The face-on shape of the thick part of the
bar is rectangular-like in the inner parts, but not in its outermost
parts where it is more oval-like. In the lower panel, the
outermost B/P parts have a clearly rectangular outline, so that this
feature could be called boxy.  
 
In the face-on view the outer part of the bar
is less thin than in the previous example, and near its ends, it has
extensions similar to the previous example, i.e. shaped as ansae. 

\section{Comparisons with observations}
\label{sec:observations}

\subsection{Morphology and photometry}
\label{subsec:MP}

As already discussed in the previous section, a face-on view is not
favourable to viewing the thick part of the bar, while edge-on views
are not favourable for the thin part of it. The best compromise comes
from intermediate, 
but close to edge-on cases. This was first noted by
\cite{Bettoni.Galletta.94} for NGC 4442 which has an inclination angle
$i$=72$^\circ$, \cite{Quillen.KFD.97} for NGC 7582 with $i$=65$^\circ$ 
and \cite{Athanassoula.Beaton.06} for M31 with $i$=77$^\circ$. In
this last paper Athanassoula \& Beaton viewed N-body simulations from
different viewing angles to compare with the near infrared (NIR) observations of
\cite{Beaton.MGSCGPAB.07}. The B/P is easily recognised, while the
outer thin part of the bar contributes two `elongations' which appear
offset from major axis of the B/P isodensities. For the inclination of
M31, this offset is best seen when
the angle between the bar and the galaxy major axes is between
20$^\circ$ and 50$^\circ$, but this range could well be somewhat model
dependent. \cite{Erwin.Debattista.13} extended this study to smaller
inclinations and showed that the B/P feature can be detected even at
inclinations as low as 40$^\circ$, although the range of bar position
angles for which the `elongations' (here called `spurs') are clearly
visible is considerably diminished. Using a sample of
78 nearby early type barred galaxies with inclinations less than
65$^\circ$ they showed that the extent of the thick part of the bar is
between 0.4 and 3.8 kpc and the relative extent compared to that of
the total bar is 0.38 $\pm$ 0.08. 

It is possible to obtain information on both the bar and the boxy/peanut
in edge-on galaxies by using photometric profiles from strips parallel to 
the major axis (i.e. the projected equatorial plane). The signature of
the bar on the profile along the major axis is a ledge followed by a
sharp drop of the intensity. The distance of the drop from the centre of
the galaxy gives the length of the bar projected on
the plane of the sky. Similarly, profiles from strips offset from the major axis
give the projected length of the B/P feature. This technique has been
widely used 
\citep[e.g.][]{Wakamatsu.Hamabe.84, Dettmar.Barteldrees.90,
  Donofrio.CMZB.99}. 

\cite{Lutticke.DP.00a} analysed a sample of  about 1350 edge-on disc
galaxies and found that about 45\% of all bulges are B/P
shaped. In a sequel paper \citep{Lutticke.DP.00b} they analysed 
photometry of 60 edge-on galaxies in the NIR to minimise the
effect of dust 
and concentrate on the old stellar population. They found a
correlation between prominent B/P bulges and strong bar signatures,
which they interpret as a dependence of the boxiness on the bar
strength, as was later
confirmed by simulations (Sect.~\ref{subsec:general}). They also
give the ratio of the bar extent to that of the B/P. Unfortunately, they
measured the bar length up to the end of the density drop, which
systematically overestimates the bar length and makes comparisons with
other works difficult.

\begin{figure*}
\centering
\includegraphics[scale=0.45, angle=0.0]{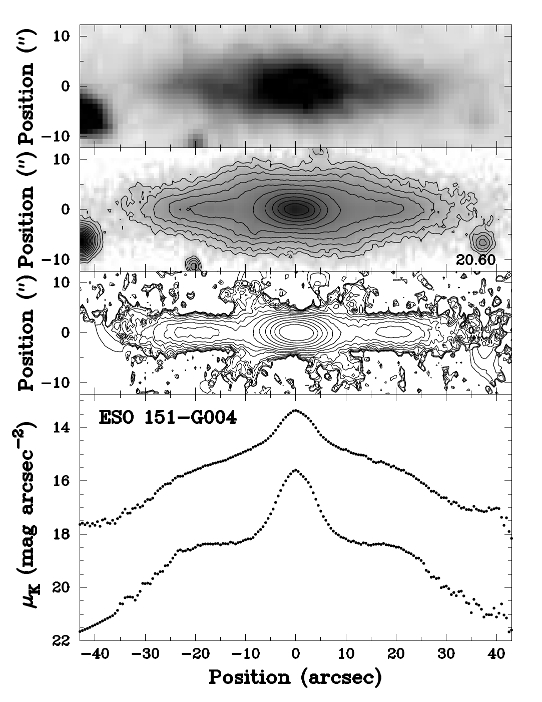}
\caption{Images and surface brightness profiles of ESO 151-G004.
 From top to bottom we have a SDSS image of the galaxy, 
 a $K$n-band image, a median-filtered $K$n-band image, and 
 major-axis (fainter) and summed (brighter) surface brightness
 profiles, all spatially registered. (Panel from figure 1  
of Bureau et al., {\it K-band observations of boxy bulges -
I. Morphology and surface brightness profiles}, 2006, MNRAS, 370, 753). 
}
\label{fig:Bureau+-obs}
\end{figure*}

\cite{Bureau.AADBF.06} analysed the structure and morphology of 30
edge-on galaxies using $K$n-band images. With the help of
unsharp masking they showed that 
galaxies with a B/P structure have more complex morphology than those
without it, revealing centred or off-centred X shapes and secondary
maxima. These are an ``essentially
near-perfect match'' to the unsharp masks of images of N-body
simulations given by \cite{Athanassoula.05a} (Sect.~\ref{sec:sims}).    
ESO 151-G004 (Fig.~\ref{fig:Bureau+-obs}) is a good example
of an off-centred X shape and NGC 1381 of a centred one (see Fig. 1 of
\citealt{Bureau.AADBF.06}). A large fraction of the galaxies have also
two secondary maxima, one on each side of the centre, similar to and at
similar locations as the unsharp masked simulated galaxies 
\citep{Athanassoula.05a}. More unsharp masked galaxies, agreeing
equally well with simulations, can be found in
\cite{Aronica.ABBDVP.03} and \cite{Patsis.Xilouris.06}.

Bureau et al. also compared two types of
surface brightness radial profiles, one from the major axis surface
brightness (lower curve in the bottom panel of
Fig.~\ref{fig:Bureau+-obs}) and the other by summing the data
vertically until
the noise level of the image was reached (upper, brighter curve in the
same panel and figure). The difference between the two argues 
that the vertical scale length varies with radius. Axisymmetric
disc galaxies generally have only two major components\footnote{Since
  here I
  concentrate on the structure of the inner parts, I do not
  discuss outer breaks and the discs beyond them.}: the disc and the classical
bulge. Yet galaxies with a B/P structure have three or four
characteristic regions. From the innermost to the outermost,
the first region has a very steep profile
and is generally associated with the bulge(s). Then follows a shallow
or even flat region which is associated with the bar and in some cases
links outwards to the disc component. In other cases, in between
this shallow component and the disc there is a steep drop,
associated with the corresponding features seen at the end of face-on bars
both in observed and in simulated galaxies. All these
features show up better on the cuts along the major axis than on
vertically summed ones, as expected.

\subsection{Kinematics}
\label{subsec:kinematics}

Position velocity diagrams (hereafter PVDs) obtained from emission line,
long slit spectra of galaxies with B/P bulges 
\citep{Kuijken.Merrifield.95, Merrifield.Kuijken.99, 
Bureau.Freeman.99} show a number of interesting features, of which the
most important has 
the form of a tilted X with one near-vertical branch 
and the other at an angle, and a clear gap between the two. There is
also material in the 
so-called forbidden quadrants. These features were already
linked to bars by \cite{Kuijken.Merrifield.95}.  

\cite{Bureau.Athanassoula.99} made model PVDs from the planar periodic orbits
in a barred galaxy model. Although this approach is too crude to
reproduce, even approximately, observed PVDs it can give valuable
insight. There are clear signatures of the $x_1$ and
the $x_2$ families. The latter is near-vertical in the PVD space,
while the former is at an angle to it. Furthermore there is signal
in the forbidden quadrants, resulting from the elongated shape of the orbits. 
To actually model emission line PVDs, \cite{Athanassoula.Bureau.99}  
used the gas flow simulations of \cite{Athanassoula.92b} viewing them
edge-on. They found that the shocks along the leading edges of the bar
and the resulting inflow lead to the characteristic gap seen in
observed PVDs. This gap thus reliably indicates the presence of a
bar and the existence of an ILR. It also 
sets strong constraints on the orientation of the bar with respect
to the line of sight.

\cite{Chung.Bureau.04} made long-slit absorption line kinematic
observations along the major axis of the 30 galaxies of the
\cite{Bureau.Freeman.99} sample. They used Gauss-Hermite series up to
fourth order and 
obtained the integrated light, the mean stellar velocity $V$, the 
velocity dispersion $\sigma$ and the third and fourth order moments 
$h_3$ and $h_4$. \cite{Bureau.Athanassoula.05} used the same
techniques and, in as much as possible, also the same software to
`observe' N-body simulations from an edge-on perspective.
They found similar signatures in these
profiles, namely i) a rotation curve with characteristic double hump,
ii) an $h_3$ that correlates with $V$ over most of the bar extent and      
iii) a velocity dispersion with a central peak which in the centre-most
region may be flat or have a relatively shallow minimum. At
intermediate radii $\sigma$ has  
a plateau, which may end on either side by a shallow maximum before
a steep drop (see also AM02).

The work described so far has only considered 1D velocity information on
a slit along the major axis. Obtaining information beyond this for NGC 4565,
\cite{Kormendy.Illingworth.82} made a very interesting finding, namely
that, within the bulge, the rotational velocity changes very little
with height, which was dubbed `cylindrical rotation'. This was confirmed
for other galaxies by many other studies
\cite[e.g.][and references therein]{Bettoni.Galletta.94, Fisher.IF.94,
  FalconB+.04, Williams.ZBKMZK.11}. From the simulation side, a very
spectacular cylindrical rotation was found by
AM02 for a strongly barred galaxy viewed
side-on. Nevertheless, although there may be some rough relation
between bar strength and cylindrical rotation, it is far from being a
clear correlation, as was found from the observational side by \cite{
  Williams.ZBKMZK.11} and from the simulations by IA15.
 
IA15 extended previous work, by including the second dimension and by
using Voronoi binning
and the software of Cappellari \citep{Cappellari.Copin.03}.  
They also used simulations including gas, star formation, feedback and
cooling, partly from \cite{Athanassoula.MR.13}. They recover the results of
\cite{Bureau.Athanassoula.05} and also find peanut related
signatures (elongated wings of large $h_3$ values and X-shaped regions
of deep $h_4$ minima) roughly in an area covering the peanut.  

When viewed end-on, bars can be mistaken for classical
bulges (e.g. AM02). This holds also for small departures
of the bar major axis from the line of sight, not exceeding 10$^{\circ}$
\citep{Athanassoula.05a}. IA15 investigated
the case where both a classical bulge and an end-on viewed bar are
present and note that the existence of the bar can be seen in the
kinematics, although its signatures are considerably weaker than in
the absence of the classical bulge, erroneously hinting
to a much weaker bar than actually present. 

Similar work, but for face-on views showed that the kinematic signature of
a face-on peanut is two minima, one on either side of the centre
\citep{Debattista.CMM.04, MendezAbreu.CDDAP.08}. These results were
recovered also by  
\cite{Iannuzzi.Athanassoula.15}, who also examined the
kinematic signatures of the second bucklings. These are much deeper
than the corresponding ones of the first buckling and could therefore
be easier to observe. Furthermore, the second buckling lasts longer
than the first one, which means there would be a higher probability to
observe it. 

Note that a few large integral-field
spectroscopic surveys of nearby galaxies are already available, and
many more are starting. Such data, particularly from large telescopes,
can provide
important new information to further our understanding of bars.

\section{Barlenses}
\label{sec:barlens}

As mentioned in the Introduction, barlenses were introduced as
separate components only very recently, so very little theoretical
work on these structures has so far been
made. \cite{Athanassoula.MR.13} ran a number of high resolution
simulations including gas and its physics (star formation, feedback
and cooling) and found very
realistic morphologies (see their figures 4 and 5 for face-on views). In
particular, the inner parts of the bars showed structures whose
morphology is very reminiscent of barlenses (see also
Figs.~\ref{fig:four-views-strong} and \ref{fig:two-views-weak} here). To
substantiate this visual impression and to understand the origin of
these structures \citet[][hereafter ALSB]{Athanassoula.LSB.14} created
fits images from 
the snapshots of these runs and analysed them using the same 
procedures and software as those used for the analysis of real galaxy
images \citep[e.g.][]{Laurikainen.SBKC.10}.

\begin{figure*}
\centering
\includegraphics[scale=0.71, angle=0.0]{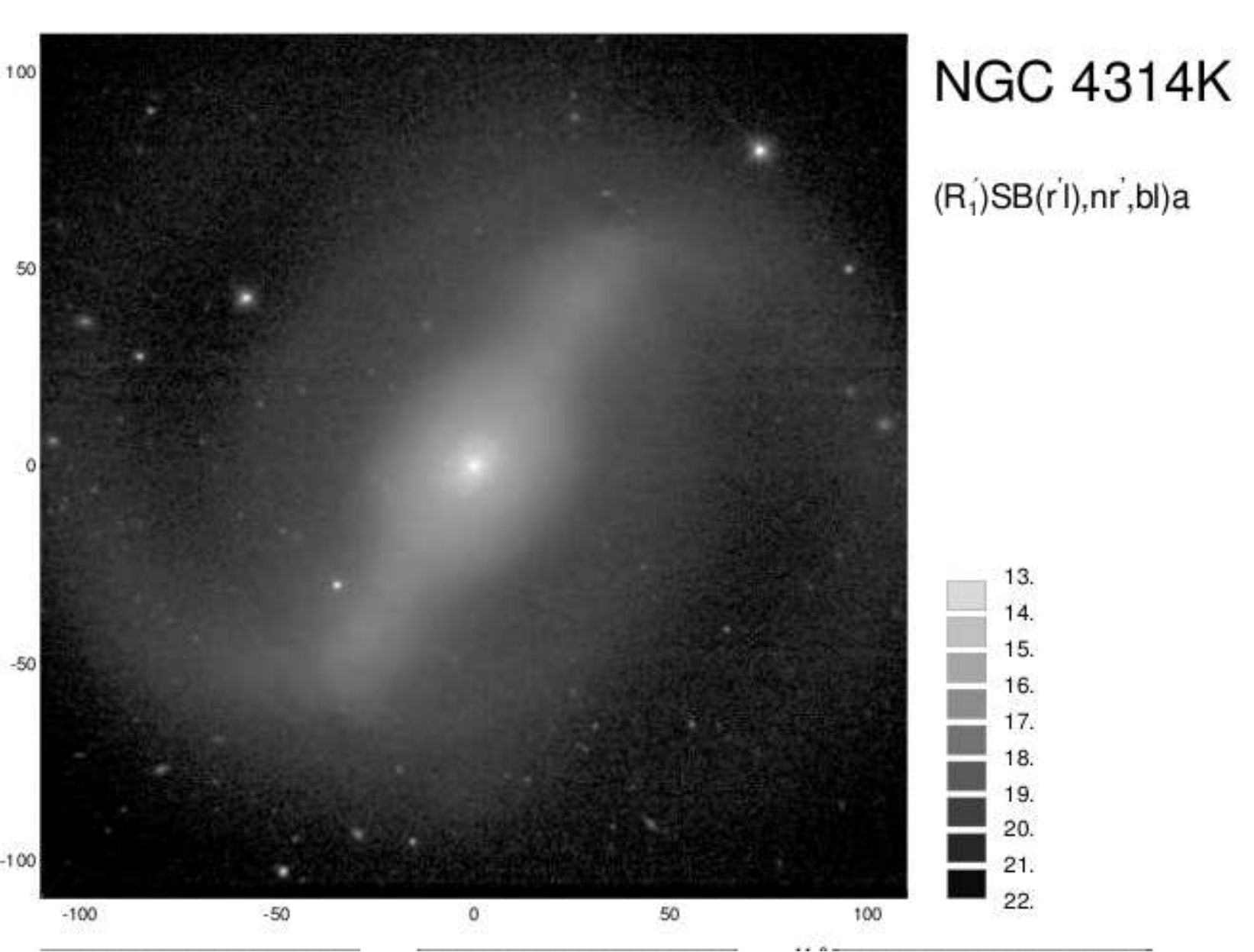}
\caption{NIRS0S image of NGC 4314. The two components of the bar are
    clearly discernible. The inner component is shorter and fatter and
    the outer one longer and more elongated (reproduced from
    \citealt{Laurikainen.SBK.11}, {\it Near-infrared atlas of S0-Sa
      galaxies (NIRS0S)}, 2011, MNRAS, 418, 1452).
}
\label{fig:N4314}
\end{figure*}

The comparison to observations started by a visual morphological
assessment, which allowed ALSB to make comparisons of observed
and simulated galaxy images. The simulated radial
projected density profiles along the bar major and minor axes are 
very similar to those found from observations. Ellipse fits and
decompositions of the simulated galaxy images allowed
further comparisons. The results are in very good agreement with those
found from observed galaxies and showed that the structures
found in the simulations can indeed be called barlenses.

An understanding of the nature of the barlens components was now
possible from an analysis of the simulations. By viewing snapshots
from many different angles, ALSB showed that {\it barlenses are the
vertically thick part of the bar viewed face-on, i.e. a barlens and a
boxy/peanut/X bulge are the same component, but simply viewed from a
different viewing angle}.

ALSB, furthermore, came up with a rule of thumb to estimate the extent of the
thick part of the bar along the bar major axis, simply from the shape
of the isophotes. Based on this, it is possible to estimate
galactic potentials more accurately \citep{Fragkoudi.ABI.15}, leading
to improvements both in orbital and gas flow calculations.  
ALSB also found from the simulation data
correlations between the bar strength and barlens related quantities.
These were confirmed by observations in \cite{Laurikainen.SABHE.14}.    

A further interesting point 
is that the barlens component can be, in some cases, mistaken for a
classical bulge (ALSB). Thus
the fraction of disc galaxies with no classical bulge is presumably
larger than what is actually acknowledged. Furthermore, in
decompositions where the barlens is taken into account as a separate
component, the mass of the classical bulge relative to the total $(B/T)$
is considerably smaller
than that found when a single component is used to model the bar. 
\cite{Laurikainen.SABHE.14} find for their sample that $<B/T>$=0.1
when a bl component is included in the decomposition, compared to 
$<B/T>$=0.35 obtained from similar decompositions when the barlens
component is omitted. 

\section{Nomenclature}
\label{sec:names}

This situation in which the same object, namely the thick part of the
bar, is known by several different names depending on the viewing
angle is not very satisfactory. 
It comes from the fact that observations preceded theory and the
thick part of the bar was observed from different angles well before 
N-body simulations and orbital structure studies established the 3D
shape of bars.

The situation is further complicated by the fact that the thick part
is rightfully called a bulge by the two most widely used definitions of a
bulge. According to the first definition, a bulge has a smooth light
distribution that swells out of the central part of a disc viewed edge-on.
B/P/Xs clearly fulfil this definition. The second 
definition of bulges is based on radial photometric 
profiles. Here the bulge is identified as the additional light in the
central part of the 
disc, above the exponential profile fitting the remaining
(non-central) part.
Barlenses clearly fulfil this definition
\citep[ALSB][]{Laurikainen.SABHE.14}.   

Thus B/P/X/bl objects (i.e. the inner thick part of the bar)
deserve to be called `bulges' with both current definitions. This is
unsatisfactory and calls
for a change of the definition of a `bulge' so as to include
kinematics. Bulges should be defined as objects that protrude out of the
galactic disc in edge-on galaxies, AND contribute to the additional light in the
inner parts of the radial photometric profile, above the disc
exponential profile, AND are clearly more pressure than rotationally
supported (as measured e.g. by their $<V>/\sigma$ value). Under this
definition, only what is now known as classical bulges
\citep{Kormendy.Kennicutt.04, Athanassoula.05a} would qualify as
bulges. B/P/X/bl objects could then rightfully be called the
thick part of the bar, while discy pseudo-bulges  
\citep{Kormendy.Kennicutt.04, Athanassoula.05a, Erwin.08} could then be
called inner discs. This suggestion deserves some consideration, since the 
nomenclature problem in this subject is becoming quite acute.  

\section{Summary}
\label{sec:summary}

Shortly after their formation, bars become vertically unstable. At
that point they may, or may not buckle out of the equatorial plane of
the  galaxy. Following this possible asymmetric stage, or directly
after the onset of the instability, the 
inner parts of the bar thicken considerably and take the shape of a
box. At the same time
the bar weakens. Subsequently in most cases the bar
amplitude starts growing again, although many cases have been found
where it stays roughly constant. The thickness of its inner part also
increases with time and its shape can evolve to peanut- or X- like.

By singling out the particles that constitute the bar in the face-on
view and then viewing the volume they occupy, it becomes clear that  
the bar has a very complex and interesting three dimensional shape. It
has a vertically thick inner part and a thin outer part. Seen face-on the
inner part is elongated along the bar; seen edge-on
it has a box, or peanut, or `X' shape. This  global bar geometry has
clear signatures when seen from different viewing angles. 

Orbital structure theory has provided the families of 3D orbits that
can constitute the backbone of this component. 
Their extent along the bar major axis 
is always smaller than that of the bar, but it varies from one family 
to another as do their
vertical height and shape. Thus the $x_1v_1$ family provides the
building blocks that are shortest along the bar major axis, vertically
thickest and face-on most  
elongated, while higher order families have orbits which are 
relatively more extended along the bar major axis, vertically thinner
and less elongated.  

Simulations, orbital structure results and observations have been
extensively inter-compared and an excellent agreement has been
found. It is clear that all three are describing the same objects.

The vertically thick part of the bar is known by different
names. Viewed edge-on it is usually referred to as boxy, peanut, or
X-shaped bulge. Viewed face-on it is known as the barlens component.  

\begin{acknowledgement}

I thank Albert Bosma for many stimulating discussions and the editors
for inviting me to write this review. I acknowledge financial support
from the People Programme (Marie Curie Actions) of the European
Union's Seventh Framework Programme FP7/2007-2013/ under REA
grant agreement number PITN-GA-2011-289313 to the DAGAL network.  
I also acknowledge financial support from the CNES 
(Centre National d'Etudes Spatiales -
France) and from the ``Programme National de
Cosmologie et Galaxies" (PNCG) of CNRS/INSU, France,
and HPC resources from GENCI- TGCC/CINES
(Grants x2013047098 and x2014047098).

\end{acknowledgement}
\end{document}